\documentstyle[12pt]{article}

\begin{document}

\setlength{\oddsidemargin}{0 cm}
\setlength{\evensidemargin}{0 cm}
\setlength{\topmargin}{0 cm}
\setlength{\textheight}{22cm}
\setlength{\textwidth}{16cm}
\setlength{\baselineskip}{0.30 in}



\begin{flushright}
    INS-Rep.-1047 \\
\end{flushright}

\begin{center}

\begin{Large}
{\bf The topological string associated with
a simple singularity of type $D$}
\end{Large}

\vspace{25pt}

\noindent
  Toshio NAKATSU

\vspace{18pt}

\begin{small}
  {\it Institute for Nuclear Study, University of Tokyo} \\
  {\it Midori-cho, Tokyo 188, Japan} \\
\end{small}

\vspace{25pt}

\underline{ABSTRACT}

\end{center}

\vspace{10pt}

\begin{small}
              The partition function of  $D_{N+1}$ topological string,
the coupled system of topological gravity and
$D_{N+1}$ topological minimal matter ,
is proposed in the framework of
KP hierarchy.  It is specified by the elements of
$GL(\infty)$ which constitute the deformed family
from the $A_{2N-1}$ topological string.
Its dispersionless limit is investigated from the view of both
dispersionless KP hierarchy and singularity theory.  In particular
the free energy restricted on the small phase
space coincides with that for the topological Landau-Ginzburg
model of type $D_{N+1}$.

\end{small}

\newpage


        Recently much attention has been paid to the understanding
of topological string theory.
In particular the topological string of type $A$ \cite{Witten}
has been studied from several perspectives.
It's partition function
$e^{\mbox{$\cal F$}_{A}}$ is defined as the $\tau$ function
of Kadomtsev-Petriashivil (KP) hierarchy
\cite{Kar-IZ,AM} which satisfies \cite{NKNT} the genus expansion
of the form :
$\mbox{$\cal F$}_{A}= \sum_{g =0}^{\infty}\mbox{$\cal F$}_{A,g}$.
$\mbox{$\cal F$}_{A,g}$ is the free energy on the 2-surface of
genus $g$. By taking an appropriate ``small phase space'' $T_s$
of the space of KP times, the quantity
$\mbox{$\cal F$}_{A}\left|_{T_s}\right. $ is shown  \cite{NKNT,DVV}
to be deeply related with the versal deformation of a simple
singularity of type $A$.

                  In spite of this  progress of our understanding
on the topological string of type $A$,
the other noncritical topological strings associated with simple
singularities have not been known well.
In this letter we try to describe the partition function of the
topological string which is associated with a simple singularity of
type $D$ .
We begin in the section 1  by studying the deformation of topological
string of type $A$. The dispersionless limit of this deformation
is investigated in the section 2, which is shown to be deeply
related with the versal deformation of a simple singularity of
type $D$. In the last section we give some discussion and propose
the partition function of topological string of type $D$.


\section{Deformation of $A_{2N-1}$ topological string}

               The $A_{2N-1}$  topological string is
the coupled system of topological gravity and
$A_{2N-1}$ topological minimal matter
\cite{Witten}.
Since the partition function
$e^{\mbox{$\cal{F}$}_{A_{2N-1}}}$,
which  has been studied from the view of matrix integrals,
is a $\tau$ function of KP hierarchy \cite{Kar-IZ},
it can be realized in terms of free fermion system,
$\psi(z)=\sum_l \psi_l z^{-l}$,
$\psi^*(z)=\sum_l \psi^*_l z^{-l-1}$ \cite{DKJM}
\footnote{$\{ \psi_l,\psi^*_m \} =\delta_{l+m,0}$.}:
\begin{equation}
\exp \left\{
        \mbox{$\cal{F}$}_{A_{2N-1}}
                       (t ; \hbar) \right \}
=
\langle 0 |
    \exp \left\{
       \sum_{ k \not\equiv  0 ~mod2N ,~ k \geq 1}
            \frac{t_k}{\hbar}J_k
                                 \right\} g_0
| 0 \rangle ,
\label{FA}
\end{equation}
where the KP time
$t_{2Nm+n}$ ($m \geq 0,1 \leq n \leq 2N-1$) is the parameter coupled with
$\sigma_m(\mbox{$\cal{O}$}_n)$,
the m-th gravitational descendants of primary field
$\mbox{$\cal{O}$}_n$ ( $\mbox{$\cal{O}$}_1= P$ :
puncture operator )  and
$"\hbar"$ plays the role of cosmological constant
of this string theory.
$J(z)=\sum_lJ_lz^{-l-1}$ is a $U(1)$ current of this
fermion system, that is, $J(z)=:\psi\psi^*:(z)$.
The vacuum $\left| 0 \rangle \right.$ is introduced  by the conditions :
$\psi_k \left| 0 \rangle \right.=0$
for $k \geq 1$ ,
and $\psi^*_k \left| 0 \rangle =0 \right.$ for
$k \geq 0$ .
$g_0$ in (\ref{FA}) is an element of $GL(\infty)$
which characterizes the partition function.

                      With the above brief review of
the $A_{2N-1}$ topological string
let us study the following perturbation of this system :
\begin{eqnarray}
\exp \left\{
        \mbox{$\cal{F}$}
                       (t_{odd} ;t^{*2}; \hbar) \right\}
&=&
   \langle 0 |
         \exp \left\{
             \sum_{k \geq 1}
                    \frac{t_{2k-1}}{\hbar}J_{2k-1}
                                 \right\} g(t^{*2})
                                         | 0 \rangle, \nonumber \\
g(t^{*2})
&=&
g_0 \exp \left\{
              \frac{t^{*2}}{2 \hbar}
                          J_{-1} \right\},
\label{FD}
\end{eqnarray}
where $t^{*}$ is a deformation parameter ($g(0)=g_0$)
and we set the even KP times equal to zero, that is,
$t_{2k}\equiv 0$ for $~^{\forall}k \geq 1$.
$t_{odd}=(t_1,t_3,\cdots)$.
Notice that $\mbox{$\cal{F}$}(t_{odd};t^{*2};\hbar)$
reduces to
$\mbox{$\cal{F}$}_{A_{2N-1}}(t_{odd};\hbar)$ at $t^{*}=0$.
Let us describe $g(t^{*2})$ in (\ref{FD}), an
element of $GL(\infty)$, in some detail.
For this purpose we shall introduce an infinite dimensional vector
space
$\mbox{$\cal{V}$}
= \oplus _{n \in \mbox{$\bf Z$}} \mbox{$\bf C$}e_{-n}(x)$
on which $\hbar \partial_x$ and $x$ act as
\footnote{$e_n(x)$
has the form:
$e_n(x)= \int d\lambda ~ e^{\frac{x \lambda}{\hbar}}\lambda^{-n-1}$.}
\begin{equation}
\hbar \partial_x e_{n}(x)
=e_{n-1}(x), ~~~x e_n(x)=\hbar (n+1) e_{n+1}(x).
\label{en}
\end{equation}
Notice that the fermion modes $\psi_n$,$\psi^*_n$
and the vacuum $\left| 0 \rangle \right.$ can
be written in terms of these bases
$\left\{ e_{-n}(x) \right\}_{n \in \mbox{$\bf Z$}}$ :
\begin{eqnarray}
&&
\left| 0 \rangle \right.
         = e_{-1}(x) \wedge e_{-2}(x) \wedge \cdots,
\nonumber
\\
&&
\psi_n= e_{-n}(x) \wedge~,~~~\psi^*_n = i_{e_n(x)}.
\end{eqnarray}
With these correspondences any pseudo-differential operator
can be realized by the free fermion.
Say, for an example,
$J_k = :(\hbar \partial_x)^k:$ holds.
$\exp \left\{ \frac{t^{*2}}{2\hbar}J_{-1}\right\}\equiv
\exp \left\{ \frac{t^{*2}}{2\hbar}:(\hbar \partial_x)^{-1}:\right\}$
transforms $e_{-n}(x)$ into
\begin{equation}
e_{-n}(x) \rightarrow
\sum_{l \geq 0}
(\frac{t^{*2}}{2\hbar})^l \int d \lambda ~
    e^{\frac{x \lambda}{\hbar}}\lambda^{n-1-l},
\label{Jtrans}
\end{equation}
and the action of $g_0$ on $\cal{V}$ is \cite{AM}
\begin{eqnarray}
e_{-n}(x) \rightarrow
\phi_n(x)= \int d \lambda ~
e^{\frac{x \lambda}{\hbar}}\hat{\phi}_n(\lambda),
\label{Abase}
\end{eqnarray}
where $\hat{\phi}_n$ is given by
\begin{eqnarray}
\hat{\phi}_n(\lambda)=
\left( \frac{\lambda^{2N-1}}{\hbar} \right)^{\frac{1}{2}}
e^{-\frac{2N}{2N+1}\frac{\lambda^{2N+1}}{\hbar}}
\cdot \int dz ~ z^{n-1}
e^{\frac{1}{\hbar}
      \left\{ -\frac{z^{2N+1}}{2N+1}+\lambda^{2N}z \right\} }.
\label{Abase2}
\end{eqnarray}
The integration with respect to $z$ in (\ref{Abase2}) is performed
by the saddle point method
\footnote{By the saddle point method
$\hat{\phi}_n(\lambda)$ has the form:
$\hat{\phi}_n(\lambda)=\sum_{l \geq 0}d_l^{(n)}\lambda^{n-l(2N+1)}$}.
By combining these two
transforms of $e_{-n}(x)$ the action of $g(t^{*2})$ on
$\cal{V}$ is decribed as
\begin{equation}
e_{-n}(x) \rightarrow
\phi_n(x;t^{*2})=
    \sum_{l \geq 0}\frac{1}{l!}(\frac{t^{*2}}{2\hbar})^l
                          \phi_{n-l}(x).
\label{Dbase1}
\end{equation}
$\phi_n(x;t^{*2})$ has the following equivalent expression  :
\begin{eqnarray}
\phi_n(x;t^{*2}) &=&
 \int d \lambda ~ e^{\frac{x \lambda}{\hbar}}
        \hat{\phi}_n(\lambda ; t^{*2}),
\label{Dbase2}  \\
\hat{\phi}_n(\lambda;t^{*2}) &=&
    \left( \frac{\lambda^{2N-1}}{\hbar} \right)^{\frac{1}{2}}
              e^{-\frac{2N}{2N+1}\frac{\lambda^{2N+1}}{\hbar}}
                \cdot   \int dz ~ z^{n-1}
                      e^{\frac{1}{\hbar}
                          \left\{ -\frac{z^{2N+1}}{2N+1}
                                +\lambda^{2N}z
                                   +\frac{t^{*2}}{2}z^{-1}\right\} },
\nonumber
\end{eqnarray}
where the integration with respect to $z$ is performed
after the change of the variable
: $z \rightarrow z'=z-\lambda$ .

                         Because, for an each fixed value of $t^*$,
$e^{\mbox{$\cal{F}$}(t_{odd};t^{*2};\hbar)}$
(\ref{FD}) can be regarded as a $\tau$-function of the KP hierarchy
with even KP times setting  zero
($t_{2k}\equiv 0$),
one can introduce
$W(t^{*2};\hbar \partial_x)$,
the corresponding wave operator of
KP hierarchy, as
\begin{equation}
W(t^{*2};\hbar \partial_x)=
1+
 \sum_{l \geq 1}w_l(t^{*2};\hbar) (\hbar \partial_x)^{-l},
\label{DW}
\end{equation}
where $w_l(t^{*2};\hbar)\equiv w_l(t_{odd};t^{*2};\hbar)$
and we identify $t_1=x$.
Through the element
$g(t^{*2}) \in GL(\infty)$
one can also associate
$e^{\mbox{$\cal{F}$}(t_{odd};t^{*2};\hbar)}$
(\ref{FD}) with a point of the
Universal Grassmann manifold (UGM) \cite{SS}.
$V(t^{*2})$, the corresponding point of UGM, is given by
\begin{equation}
V(t^{*2})=
\oplus_{n \geq1}\mbox{$\bf{C}$}\phi_n(x;t^{*2}).
\label{DUGM}
\end{equation}
The relation between $V(t^{*2})$, the underlying point of UGM,
and $W(t^{*2};\hbar \partial_x)$, the wave operator of KP hierarchy,
is \cite{SN}
\begin{equation}
V(t^{*2})=
 W_0(t^{*2};\hbar \partial_x)^{-1}V_{\emptyset},
\label{Theorem}
\end{equation}
where $V_{\emptyset}$ is the point of UGM which corresponds to
the vacuum $\left| 0 \rangle \right.$,
that is, $V_{\emptyset}=\oplus_{n \geq 1}\mbox{$\bf{C}$}e_{-n}(x)$,  and
$W_0(t^{*2};\hbar \partial_x)
 =W(t^{*2};\hbar \partial_x)
                |_{t_{odd}=(x,0,\cdots)}$ is
the initial value of the wave operator.
We shall characterize the wave operator
(\ref{DW}) from the study of the point of UGM (\ref{DUGM}) .
In particular we should notice that the bases
$\left\{ \phi_n(x;t^{*2}) \right\}_{n \in \mbox{$\bf{Z}$}}$
(\ref{Dbase1}) satisfy the following set of equations:
\begin{eqnarray}
 F~\phi_n(x;t^{*2}) &=&
    \phi_{n+2N}(x;t^{*2})+\hbar(1-n)\phi_{n-1}(x;t^{*2})
                  +\frac{t^{*2}}{2}\phi_{n-2}(x;t^{*2}),
\nonumber \\
&&
\label{eq1} \\
G~\phi_n(x;t^{*2}) &=& -\phi_{n+1}(x;t^{*2}),
\label{eq2}
\end{eqnarray}
where we introduce the operators $F$ and $G$ as
\begin{equation}
F=(\hbar \partial_x)^{2N},~~
G=\frac{1}{2N}\left\{
        x-2N(\hbar \partial_x)^{2N}
              -\frac{2N-1}{2}\partial_x^{-1} \right\}
                  (\hbar \partial_x)^{-2N+1}.
\label{FG}
\end{equation}
By combining these equations with (\ref{Theorem})  we can see
\begin{eqnarray}
P_0 &=&
(\hbar \partial_x)^{2N}+x+\frac{t^{*2}}{2}(\hbar \partial_x)^{-2},
\label{string1}
\\
Q_0 &=& -\hbar \partial_x,
\label{string2}
\end{eqnarray}
where we set
$P_0=W_0(t^{*2};\hbar \partial_x)FW_0(t^{*2};\hbar\partial_x)^{-1}$ and
$Q_0=W_0(t^{*2};\hbar \partial_x)GW_0(t^{*2};\hbar\partial_x)^{-1}$.
These two operators can be also regarded as
$P_0=P|_{t_{odd}=(x,0,\cdots)}$ and
$Q_0=Q|_{t_{odd}=(x,0,\cdots)}$,
the initial values of the operators $P$ and $Q$
under the flows of odd KP times :
\begin{eqnarray}
P &=&
L^{2N},
\label{string3} \\
Q &=&
\frac{1}{2N} \left(
   M-2NL^{2N}-\frac{2N-1}{2}\hbar L^{-1} \right)L^{-2N+1}.
\label{string4}
\end{eqnarray}
In the above descriptions we introduce
the Lax and Orlov operators of KP
hierarchy \cite{KO} as
\begin{eqnarray}
 L &=&   W(t^{*2};\hbar \partial_x)
            \hbar \partial_x
                  W(t^{*2};\hbar\partial_x)^{-1}, \nonumber \\
 M &=&  W(t^{*2};\hbar \partial_x)
            \sum_{l \geq 1}
                 (2l-1)t_{2l-1}
                   (\hbar \partial_x)^{2l-2}
                       W(t^{*2};\hbar\partial_x)^{-1}.
\label{LM}
\end{eqnarray}
Since $Q_0$ (\ref{string2}) is a differential operator
and the time evolutions of $Q$ (\ref{string4}) follow from those
of the KP hierarchy
\footnote{
$(~)_{\pm}$ are the projection operators with respect to
$\hbar \partial_x$ :
\begin{eqnarray*}
((\hbar \partial_x)^m)_+
= \left\{ \begin{array}{ll}
          (\hbar \partial_x)^m & m \geq 0 \\
                      0        & m < 0
          \end {array}
  \right. ~~~~
((\hbar \partial_x)^m)_-=(\hbar \partial_x)^m-((\hbar \partial_x)^m)_+.
\end{eqnarray*}
},
\begin{equation}
\hbar \partial_{t_{2k-1}}Q
 = \mbox{[} B_{2k-1}, Q \mbox{]}~~~~
\left( B_{2k-1}=(L^{2k-1})_+ \right) ,
\end{equation}
this property of $Q_0$ is preserved under the flows :
\begin{equation}
(Q)_-=0.
\label{string5}
\end{equation}

                    At this stage it may be convenient to give some remarks.
Let us attach the topological
weight, "$wt$" , to the parameters \cite{NKNT}:
\begin{equation}
wt(t_{k})=k-2N-1, ~~~~wt(\hbar)=-2N-1.
\label{TW1}
\end{equation}
Notice that these topological weights reflect the ghost numbers of
the observables for $A_{2N-1}$
topological string \cite{Witten}.  In particular the action
of $g_0$ (\ref{Abase}) on $\cal{V}$ preserves these weights
\footnote{One can also attach the
topological weights to $\lambda$ and $z$ in (\ref{Abase2})
as $wt(\lambda)=wt(z)=-1$.}.
We shall also introduce  the topological weight of  $t^{*}$ as
\begin{equation}
wt(t^*)=-N-1,
\label{TW2}
\end{equation}
so that the action of $g(t^{*2})$ (\ref{Dbase1}) preserves them.
Consequently, from eqs. (\ref{string1}) and
(\ref{string2}), the operators $P$ and $Q$
are guasi-homogeneous with respect to these topological
weights (\ref{TW1}) and (\ref{TW2}):
\begin{equation}
wt(P)=-2N,~~~~wt(Q)=-1,
\label{TW3}
\end{equation}
which lead to the guasi-homogeneity of $L$ and $M$ :
\begin{equation}
wt(L)=-1,~~~~wt(M)=-2N.
\label{TW4}
\end{equation}
Since one can expand the Orlov operator $M$
in terms of the Lax operator $L$ as
\footnote{
$P_l(t_{odd})$ is a Schur polynomial
with $t_{even}=0$ which generating function
is $e^{\sum_{l \geq 1}t_{2l-1}\lambda^{2l-1}}
=\sum_{l \geq 0}P_l(t_{odd})\lambda^l$,
and $\tilde{\partial}_{t_{odd}}=(
\partial_{t_1},\frac{\partial_{t_3}}{3},\cdots)$.}
\begin{equation}
M=
\sum_{l \geq 1}(2l-1)t_{2l-1}L^{2l-2}
  -\hbar \sum_{l \geq 1}l P_l(-\hbar \tilde{\partial}_{t_{odd}})
      \mbox{$\cal{F}$}(t_{odd};t^{*2};\hbar)
                          L^{-l-1},
\end{equation}
$\mbox{$\cal{F}$}(t_{odd};t^{*2};\hbar)$
is also guasi-homogeneous with respect to these topological weights:
\begin{equation}
wt(\mbox{$\cal{F}$})=0.
\label{TW5}
\end{equation}


\section{$\hbar \rightarrow 0$ limit}

             Let us study the $\hbar \rightarrow 0$ limit of
$\mbox{$\cal F$}(t_{odd};t^{*2};\hbar)$ (\ref{FD}).
The underlying integrable hierarchy is the dispersionless KP
hierarchy \cite{Kri},\cite{TT} which is a "quasi-classical" limit
of the KP hierarchy in the following substitution :
\begin{eqnarray}
\mbox{[} \hbar \partial_x,~ x \mbox{]} = \hbar
\rightarrow \{ p, x \} =1,
\end{eqnarray}
where the Poisson bracket is defined by
\begin{equation}
\{  f, g \}=
\frac{\partial f}{\partial p}\frac{\partial g}{\partial x}
-\frac{\partial f}{\partial x}\frac{\partial g}{\partial p}.
\end{equation}
The $\hbar \rightarrow 0$ limit of the Lax operator
$L$ and the Orlov operator $M$ (\ref{LM}) have the form :
\begin{eqnarray}
L \rightarrow
\mbox{$\cal L$} &=&
   p+ \sum_{l \geq 1}u_{l+1}(t_{odd};t^{*2})p^{-l} \nonumber \\
M \rightarrow
\mbox{$\cal M$} &=&
   \sum_{l \geq 1}(2l-1)t_{2l-1} \mbox{$\cal L$}^{2l-2}
      +\sum_{l \geq 1}v_{2l}(t_{odd};t^{*2})
                   \mbox{$\cal L$}^{-2l},
\label{dLM}
\end{eqnarray}
which satisfy
$\{ \mbox{$\cal L$}, \mbox{$\cal M$} \}=1$.
Notice that $v_{2l}$ in (\ref{dLM}) is given by
\begin{equation}
v_{2l}=\frac{\partial \mbox{$\cal F$}_0}{\partial t_{2l-1}},
\end{equation}
where $\mbox{$\cal F$}_0$ is the leading coefficient of the expansion
of $\cal F$ (\ref{FD}) in terms of $\hbar$ \cite{TT}:
\begin{equation}
\mbox{$\cal F$}(t_{odd};t^{*2};\hbar)
  = \hbar^{-2} \left\{
       \mbox{$\cal F$}_0(t_{odd};t^{*2})
           + \mbox{\cal o}(\hbar) \right\}.
\label{F0}
\end{equation}
One can also say that
$e^{\mbox{$\cal F$}_0}$ is a $\tau$ function of the dispersionless
KP hierarchy.

                    We shall begin by studying the $\hbar \rightarrow 0$
limit of the operator $P$ (\ref{string3}). Obviously
$\cal P$, the dispersionless limit of $P$, is given by
$\cal P$$ = $$\mbox{$\cal L$}^{2N}$.The initial value of $\cal P$ is the
dispersionless limit of the eq. (\ref{string1}) :
\begin{equation}
\mbox{$\cal P$}|_{t_{odd}=(x,0,\cdots)}
= p^{2N}+x + \frac{t^{*2}}{2}p^{-2}.
\label{initial}
\end{equation}
The time evolutions of $\cal P$ follow from those of the dispersionless
KP hierarchy
\footnote{ At the dispersionless limit
$(~)_{\pm}$ become the projection operators with respect to
$p$ :
\begin{eqnarray*}
(p^m)_+
= \left\{ \begin{array}{ll}
          p^m & m \geq 0 \\
          0        & m < 0
          \end {array}
  \right. ~~~~
(p^m)_-=p^m-(p^m)_+.
\end{eqnarray*}
} :
\begin{equation}
\frac{\partial \mbox{$\cal P$}}{\partial t_{2l-1}}
=
\{ \mbox{$\cal B$}_{2l-1},\mbox{$\cal P$} \},~~~~~
(~\mbox{$\cal B$}_{2l-1}=(\mbox{$\cal L$}^{2l-1})_+~)
\end{equation}
Notice that $\cal P$ is quasi-homogeneous with respect to the topological
weights ($wt(\mbox{$\cal P$})=-2N$). Then, by considering
these time evolutions
with the initial value (\ref{initial}), we can conclude that
$\cal P$ has the following form :
\begin{eqnarray}
\mbox{$\cal P$}
=p^{2N}+\sum_{l=1}^{N}a_l(t_{odd};t^{*2})p^{2N-2l}
           +a_*(t_{odd};t^{*2})p^{-2}.
\label{dP}
\end{eqnarray}

       Nextly we shall look at the dispersionless limit of the operator
$Q$ (\ref{string4}). It becomes
\begin{equation}
\mbox{$\cal Q$}
=\frac{1}{2N}
   \left( \mbox{$\cal M$}-2N \mbox{$\cal L$}^{2N} \right)
              \mbox{$\cal L$}^{-2N+1}.
\label{dQ}
\end{equation}
In particular the $\hbar \rightarrow 0$ limit of the condition
(\ref{string5}) is
\begin{eqnarray}
(\mbox{$\cal Q$})_- =0,
\label{dstring}
\end{eqnarray}
which is equivalent to
\begin{eqnarray}
\mbox{res} \{
  ( \mbox{$\cal Q$})_-  \mbox{$\cal L$}^k
         \frac{\partial \mbox{$\cal L$}}{\partial p}dp \}=0
{}~~~~~~~~\mbox{for}~~~^{\forall} k \geq 0.
\label{dstring2}
\end{eqnarray}
One can rephrase the condition (\ref{dstring2}) into that on
$\mbox{$\cal F$}_0$.
By substituting the expansion of $\cal M$ (\ref{dLM}) with respect to
$\cal L$ into $\cal Q$ (\ref{dQ}) and then utilizing the formula
\cite{TT}
\begin{eqnarray}
\mbox{res}
   \{\mbox{$\cal L$}^{2m+1}
       \frac{\partial \mbox{$\cal B$}_{2m+1}}{\partial p} dp \}
=\frac{\partial^2 \mbox{$\cal F$}_0}
            {\partial t_{2m+1} \partial t_{2n+1}},
\end{eqnarray}
the condition (\ref{dstring2}) turns to
\begin{eqnarray}
\frac{\partial C_0(t_{odd};t^{*2})}{\partial t_{2k+1}}  = 0 ~~~~~~~~
\mbox{for}~~~^{\forall} k \geq 0,
\label{Vir0}
\end{eqnarray}
where
\begin{eqnarray}
C_0(t_{odd};t^{*2}) &=&
\sum_{l \geq 1}(2l-1+2N)t_{2l-1+2N}
                     \frac{\partial \mbox{$\cal F$}_0}{\partial t_{2l-1}}
         -2N\frac{\partial \mbox{$\cal F$}_0}{\partial t_1}  \nonumber \\
&& ~~~~+\frac{1}{2}\sum_{l=1}^{N}(2l-1)(2N-2l+1)t_{2l-1}t_{2N-2l+1}.
\label{Vir1}
\end{eqnarray}
Notice that one can integrate the eq.(\ref{Vir0})
to the form, $C_0(t_{odd};t^{*2}) + f(t^{*2}) =0$, that is,
up to the ambiguity for the dependence of parameter $t^{*}$.
{}From the quasi-homogeneity  of $C_0$ with respect to
the topological weights this dependence will be restricted to
\begin{eqnarray}
C_0(t_{odd};t^{*2}) + \alpha t^{*2} =0,
\label{Vir2}
\end{eqnarray}
where "$\alpha$" is a constant which still remains undetermined.
We shall determine the value of $\alpha$. Notice that,
by cosidering the eq. (\ref{Vir0}) for the case of
$k=N$ and then restricting it on the initial time,
we can obtain
\begin{eqnarray}
\frac{\partial \mbox{$\cal F$}_0}
          {\partial t_1}
                \left|_{t_{odd}=(x,0,\cdots)} \right.
&=& \frac{2N}{2N+1}
       \mbox{res}\{\mbox{$\cal L$}^{2N+1}dp \}
                \left|_{t_{odd}=(x,0,\cdots)} \right. \nonumber \\
&=& \frac{1}{2}t^{*2}.
\label{const2}
\end{eqnarray}
The last equality in (\ref{const2}) is the result of the direct
calculation based on the initial condition (\ref{initial}).
One can also obtain the following relation by restricting the eq.
(\ref{Vir2}) on the initial time :
\begin{eqnarray}
\frac{\partial \mbox{$\cal F$}_0}
          {\partial t_1}
                \left|_{t_{odd}=(x,0,\cdots)} \right.
=\frac{\alpha}{2N}t^{*2}.
\end{eqnarray}
By comparing these two equations we obtain $\alpha = N $.
Hence the condition $\left( \mbox{$\cal Q$} \right)_-=0$
gives rise to the equation :
\begin{eqnarray}
C_0(t_{odd};t^{*2}) + N t^{*2} =0.
\label{Vir3}
\end{eqnarray}

              We shall introduce $T_s^{KP}$, the "small phase space" of
odd KP times as
$T_s^{KP} \ni t_{odd}=(t_1,t_3, \cdots)
\Leftrightarrow t_{2N+2k+1}=0~~$ for $~~^{\forall} k \geq 0$.
On this small phase space the following equalities hold :
For $1 \leq  k \leq N$
\begin{eqnarray}
t_{2N-2k+1}
&=&
\frac{2N}{(2N-2k+1)(2k-1)}
\mbox{res}\{
\mbox{$\cal L$}^{2k-1}dp \} \left|_{T_s^{KP}} \right. ,
\label{sKPtime} \\
\frac{\partial \mbox{$\cal F$}_0}{\partial t_{2k-1}}
\left|_{T_s^{KP}} \right.
&=&
\frac{2N}{2N+2k-1}\mbox{res}\{
\mbox{$\cal L$}^{2N+2k-1} dp \} \left|_{T^{KP}_s} \right. .
\label{sfree}
\end{eqnarray}
Though these relations are
the direct consequences of the equation (\ref{Vir3}),
they are also deeply related with the singularity theory \cite{Saito}.
In order to explain this point,
let us consider the versal deformation of a simple singularity
of type $D_{N+1}$ :
\begin{eqnarray}
h(y,z)=y^N+yz^2+\sum_{l=1}^{N}a_ly^{N-l}+bz,
\end{eqnarray}
where $(a_1,a_2,\cdots,a_N,b)\in \mbox{$\bf C$}^{\otimes N+1}$ are the
deformation parameters.The flat coordinates
$s=(s_1,s_2,\cdots,s_N,s^*)$ of this parameter space are given  \cite{NSSY}
by
\begin{eqnarray}
s_k =
\frac{2N}{2k-1}
\sum_{\alpha_1, \cdots, \alpha_N}
\left( \begin{array}{c}
   \frac{2N-1}{N} \\ \sum_{l=1}^{N} \alpha_l
      \end{array}   \right)
\frac{(\sum_{l=1}^{N}\alpha_l)!}{\prod_{l=1}^{N}\alpha_l!}
a_1^{\alpha_1}\cdots a_{N}^{\alpha_N} , ~~~~
s^*=b,
\end{eqnarray}
where the sum is performed with respect to $(\alpha_1,\cdots, \alpha_N)
\in \left( \mbox{$\bf Z$}_{\geq 0} \right)^{\otimes N}$ which satisfies
$\sum_{l=1}^{N}l \alpha_l = k$ .
Notice that the above expression for the flat coordinates can
be simplified  \cite{DVV} as
\begin{eqnarray}
s_k=\frac{2N}{2k-1}\mbox{res}\{
   H^{\frac{2k-1}{2N}}(p)dp \},~~~
\frac{s^{*2}}{2}= \mbox{res}\{pH(p)dp\},
\label{sflat}
\end{eqnarray}
where we introduce the potential function $H(p)$ as
\begin{eqnarray}
H(p)=p^{2N}+
      \sum_{l=1}^{N}a_lp^{2N-2l}+
           \frac{b^2}{2}p^{-2}.
\label{H}
\end{eqnarray}
It is important to note that,
with the identification of the potential $H$ (\ref{H})
with $\cal P$ (\ref{dP}), the eqs. (\ref{sKPtime}) and (\ref{sflat})
tell us
\begin{eqnarray}
s_k=(2N-2k+1)t_{2N-2k+1} ~~~~~\mbox{on}~~T_s^{KP}.
\end{eqnarray}
What about the flat coordinate $s^*$ ?
One can write
$a_*(t_{odd};t^{*2})$ ,
the coefficient of  $p^{-2}$ in $\cal P$ (\ref{dP}),  as
\footnote{ We use the formula \cite{TT} :
$p= \mbox{$\cal L$} - \sum_{l \geq 1}\frac{1}{l}
\mbox{res}\{ \mbox{$\cal L$}^{l}dp \} \mbox{$\cal L$}^{-l} $.
}
\begin{eqnarray}
a_*(t_{odd};t^{*2}) &=& \mbox{res}\{p \mbox{$\cal L$}^{2N}dp \}
\nonumber \\
&=&
\frac{2N}{2N+1} \mbox{res}\{ \mbox{$\cal L$}^{2N+1}dp \}
\nonumber \\
&&~~~~~
-\sum_{l=1}^{N}\frac{1}{2l-1}
       \mbox{res}\{\mbox{$\cal L$}^{2l-1}dp \}
              \mbox{res}\{\mbox{$\cal L$}^{2N-2l+1}dp \},
\end{eqnarray}
from which, by using the eqs.(\ref{Vir3}), (\ref{sKPtime}) and (\ref{sfree}),
one can obtain
\begin{eqnarray}
\mbox{res}\{ p\mbox{$\cal L$}^{2N}dp \} \left|_{T_s^{KP}} \right.
   = \frac{1}{2}t^{*2}.
\end{eqnarray}
Hence on the small phase
space $T^{KP}_s$  $s=(s_1,\cdots,s_N,s^*)$,
the flat coordinates of the versal
deformation of $D_{N+1}$ simple singularity,  coincide
with the KP times  and the deformation parameter,
$(t_1,\cdots,t_{2N-1},t^*)$ .


\section{Discussion}

              The eqs. (\ref{Vir3}), (\ref{sKPtime}) and (\ref{sfree})
mean that $\mbox{$\cal F$}_0(t_{odd};t^{*2}) \left|_{T_s^{KP}} \right.$
is equal to the free energy of $D_{N+1}$ topological
Landau-Ginzburg (LG) model \cite{DVV}.
The physical observables of $D_{N+1}$ topological LG model are
denoted as
$\tilde{\mbox{$\cal O$}}_k$ ($1 \leq k \leq N$) and $\mbox{$\cal O$}_*$.
The values of N=2 $U(1)$ charge are
$\frac{k-1}{N}$ and $\frac{N-1}{2N}$ respectively.
The KP times $t_{2k-1}$ ($1 \leq k \leq N$) and the deformation parameter
$t^*$ are those
coupled with these observables.
Moreover one can see that the following relation is consistent with the
condition (\ref{Vir3}) :
\begin{eqnarray}
&& \partial_{t_{2m-1}} \partial_{t_{2i-1}} \partial_{t_{2j-1}}
\mbox{$\cal F$}_0 \nonumber \\
&& =
\sum_{k=1}^{N}\frac{2m-1}{(2k-1)(2N-2k+1)}
\partial_{t_{2(m-N)-1}} \partial_{t_{2k-1}}
\mbox{$\cal F$}_0
\partial_{t_{2N-2k+1}} \partial_{t_{2i-1}} \partial_{t_{2j-1}}
\mbox{$\cal F$}_0
\nonumber \\
&& + \frac{2m-1}{N}
\partial_{t_{2(m-N)-1}} \partial_{t^*}
\mbox{$\cal F$}_0
\partial_{t^*} \partial_{t_{2i-1}} \partial_{t_{2j-1}}
\mbox{$\cal F$}_0 ,
\end{eqnarray}
where $m \geq N+1$ and  $1 \leq i,j \leq N$.
This can be regarded as the recursion relation on 2-sphere
for the coupled system of topological matter and topological gravity
\cite{DiWi}.
Thus we can also say that the KP time $t_{2Nm+2k-1}$
($m \geq 0, 1 \leq k \leq N$)
is the parameter coupled with $\sigma_m(\tilde{\mbox{$\cal O$}}_k)$,
the $m-$ th gravitational descendant of $\tilde{\mbox{$O$}}_k$.
With these observations it is very plausible that our model
gives the description of $D_{N+1}$ topological string. So we would like
to propose the following conjecture.
\newtheorem{guess}{Conjecture}
\begin{guess}
$\mbox{$\cal F$}$ in (\ref{FD}) satisfies
\begin{eqnarray}
\mbox{$\cal F$}(t_{odd};t^{*2};\hbar)
=\sum_{g=0}^{\infty}\hbar^{2g-2}\mbox{$\cal F$}_g(t_{odd};t^{*2}).
\label{Dstring}
\end{eqnarray}
where
$\mbox{$\cal F$}_g$,
the free energy of $D_{N+1}$ topological string on the 2-surface
of genus $g$, has the form :
\begin{eqnarray}
&& \mbox{$\cal F$}_g(t_{odd};t^{*2}) \nonumber \\
&&  =
\sum_{s,d_* ,\{ l_i,k_i,d_i\}}
\langle
\mbox{$\cal O$}_*^{d_*}
\prod_{i=1}^{s}\sigma_{l_i}(\tilde{\mbox{$\cal O$}}_{k_i})^{d_i}
\rangle
_{\scriptscriptstyle{\overline{\mbox{$\cal M$}}_{g, d_*+\sum_{i=1}^{s}d_i}}}
\frac{ (t^*)^{d_*} \prod _{i=1}^{s}(t_{2Nl_i+2k_i-1})^{d_i} }
{d_*! \prod_{i=1}^{s}d_i! } .  \nonumber \\
&&
\label{gDstring}
\end{eqnarray}
\end{guess}
$\langle \cdots \rangle
_{\scriptscriptstyle{\overline{\mbox{$\cal M$}}_{g,d_*+\sum_{i=1}^{s}d_i}}}
$
is the topological correlation function
\cite{Kon-Witten} evaluated on the compactified
moduli space
$\overline{\mbox{$\cal M$}}_{g, d_*+\sum_{i=1}^{s}d_i}$
of the Riemann surface $\Sigma$ of genus $g$ with $d_*+\sum_{i=1}^{s}d_i$
punctures.
Notice that, because of the quasi-homogeneity of $\cal F$ (\ref{TW5}),
one can see that the correlation function,
$\langle
\mbox{$\cal O$}_*^{d_*}
\prod_{i=1}^s \sigma_{l_i}(\tilde{\mbox{$\cal O$}}_{k_i})^{d_i}
\rangle
_{\scriptscriptstyle{\overline{\mbox{$\cal M$}}_{g, d_*+\sum_{i=1}^sd_i}}}$,
automatically vanishes except for the case :
\begin{eqnarray}
\sum_{i=1}^s d_i \left(
 l_i+\frac{k_i-1}{N}-1 \right) +d_* \left( \frac{N-1}{2N}-1 \right)
= (d_{D_{N+1}}-3)(1-g),
\end{eqnarray}
where $d_{D_{N+1}} = \frac{N-1}{N}$.
This is nothing but the ghost number conservation rule for the
coupled system of $D_{N+1}$ topological minimal matter
and topological gravity.


\end{document}